\documentclass[global,twocolumn]{svjour}

\usepackage{graphicx}
\usepackage{epsfig}
\usepackage{hyperref}
\usepackage{dcolumn}
\usepackage{amssymb}
\usepackage{amsmath}
\usepackage{color}

\newcommand{\beqa}{\begin{eqnarray}}
\newcommand{\eeqa}{\end{eqnarray}}
\newcommand{\beq}{\begin{equation}}
\newcommand{\eeq}{\end{equation}}

\newcommand{\ket}[1]{| #1\rangle}
\newcommand{\bra}[1]{\langle #1|}

\journalname{Applied Physics B}

\begin{document}
\title{Photonic quantum simulator for unbiased phase covariant cloning}

\author{Laura T. Knoll\inst{1}\inst{2} \and Ignacio H. L\'opez Grande\inst{1}\inst{2} \and Miguel A. Larotonda\inst{1}\inst{2}\inst{3}}

\institute{DEILAP, CITEDEF-CONICET, J.B. de La Salle 4397, 1603 Villa Martelli, Buenos Aires, Argentina
\and
Departamento de F\'{\i}sica, FCEyN, UBA. Ciudad Universitaria, 1428 Buenos Aires, Argentina
\and
CONICET, Argentina
\\\email{lknoll@citedef.gob.ar}}

\maketitle

\begin{abstract}
We present the results of a linear optics photonic implementation of a quantum circuit that simulates a phase covariant cloner, by using two different degrees of freedom of a single photon. 
We experimentally simulate the action of two mirrored $1\rightarrow 2$ cloners, each of them biasing the cloned states into opposite regions of the Bloch sphere. We show that by applying a random sequence of these two cloners, an eavesdropper can mitigate the amount of noise added to the original input state and therefore prepare clones with no bias but with the same individual fidelity, masking its presence in a quantum key distribution protocol. Input polarization qubit states are cloned into path qubit states of the same photon, which is identified as a potential eavesdropper in a quantum key distribution protocol. The device has the flexibility to produce mirrored versions that optimally clone states on either the northern or southern hemispheres of the Bloch sphere, as well as to simulate optimal and non-optimal cloning machines by tuning the asymmetry on each of the cloning machines. 
\end{abstract}

\section{Introduction}
\label{sec:intro}
The no-cloning theorem of quantum information is the formal statement of the fact that unknown quantum states cannot be perfectly copied \cite{wootters1982single,dieks1982communication}. Without this restriction one would be allowed to completely determine the quantum state of a system by measuring copies of it, even leaving the original system untouched! This theorem underlies the security of all Quantum Key Distribution (QKD) protocols, and has consequences at a technological level, imposing limitations on error correction techniques and on other tasks that involve transmission of information \cite{werner2001quantum,eberhard1989quantum,peres2004quantum}. 
Despite this strong restriction, several approximate cloning machines can be constructed. That is, the production of imperfect copies is indeed allowed. In general, this is achieved by coupling the system to be cloned to an auxiliary system and applying a trace-preserving, completely-positive map to the composite system \cite{buvzek1996quantum,bruss1998optimal,gisin1997optimal,fiuravsek2005highly,iblisdir2005generalised}.

Among the family of Quantum Cloning Machines (QCM), one can make a distinction between \emph{universal} cloning machines (UQCM), which copy all the states with the same fidelity $F$, regardless of the state $\ket{\psi}$ to be cloned, and \emph{state-dependent} QCM. For qubits, the optimum UQCM can achieve a cloning fidelity $F$ of $5/6=0.833$ \cite{buvzek1996quantum,bruss1998optimal,gisin1997optimal,gisin1998quantum}. UQCMs have been experimentally realized in photon stimulated emission setups \cite{lamas2002experimental,fasel2002quantum,fan2002cloning}, linear optics \cite{lemr2012experimental} and NMR systems \cite{cummins2002approximate}.
State-dependent machines can be designed to perform better for those restricted input states than a UQCM.

One of the most relevant features of quantum cloning is its usefulness for eavesdropping on QKD systems over noisy quantum channels. This quantum channel is controlled by Eve, the eavesdropper, who can perform any operation allowed by quantum mechanics. By exploiting quantum cloning, Eve can keep one of the output states and send another to the legitimate receiver, Bob. Eve's strategy is to try to get as much information as she can while producing as less disturbance in Bob's state. For this task, an \emph{optimal} quantum cloning machine is required, in the sense that for a given fidelity of the original state, the fidelities of the cloned states are maximal. Depending of the specific QKD protocol, different cloning machines can be designed for optimal fidelity. That is, a UQCM may not be optimal for the specific set of states that are involved in a particular QKD protocol, but rather a particular state-dependent QCM may perform optimally the task \cite{d2001optimal,d2003optimal}. 

The best-known state-dependent QCM is the so-called phase-covariant QCM. It can optimally clone states of the form $\ket{\psi}=\frac{1}{\sqrt{2}}(\ket{0}+e^{i\phi}\ket{1})$, that lie on the equator of the Bloch sphere. The Phase-Covariant Cloning (PCC) machine has a remarkable application in quantum cryptography, since it is used in the optimal incoherent strategy for eavesdropping on the BB84 QKD protocol \cite{fuchs1997optimal,cerf2002security,bruss2000phase} that runs with the bases of eigenstates of $\sigma_x$ and $\sigma_y$. The eavesdropper
on BB84 needs to gather information only on the four states $\frac{1}{\sqrt{2}}(\ket{0}\pm\ket{1})$ and $\frac{1}{\sqrt{2}}(\ket{0}\pm i \ket{1})$, however it can be shown that the optimal solution for cloning these four states and for cloning all the states from the equator is the same \cite{scarani2005quantum}.

The task of optimally cloning the equator of the Bloch sphere can be accomplished without ancilla \cite{niu1999two}. This situation is usually depicted as a $1\rightarrow 2$ cloning; a single qubit that interacts with another qubit in a blank state (Eve's qubit), which results in a transformation on a two qubit system that makes two imperfect copies of the first qubit. The asymmetric PCC transformation on Bob and Eve's qubit can be expressed as
\begin{equation}
\begin{aligned}
	&|0\rangle_B|0\rangle_E\rightarrow|0\rangle_B|0\rangle_E\\
	&|1\rangle_B|0\rangle_E\rightarrow\sqrt{1-q}|1\rangle_B|0\rangle_E+\sqrt{q}|0\rangle_B|1\rangle_E
	\label{eq:PCC}
\end{aligned}
\end{equation}
where $0\leq q\leq1$ controls the asymmetry of the cloning operation. Once applied to states on the equator, the reduced density matrices $\rho_B$ and $\rho_E$ can be obtained and the fidelities of the clones can be calculated as $F_B=\bra{\psi}\rho_B\ket{\psi}=\frac{1}{2}(1+\sqrt{1-q})$ and $F_E=\bra{\psi}\rho_E\ket{\psi}=\frac{1}{2}(1+\sqrt{q})$. Both fidelities are independent of the phase $\phi$ of the input state and for the symmetric case in which $q=0.5$ we obtain $F_B=F_E=0.8536$, outperforming a UQCM. Therefore, the PCC machine allows for a higher fidelity than that of the universal cloning machine for all states on the equator of the Bloch sphere. 

A relevant issue is that being not universal, a PCC machine comprises an operation that shrinks the Bloch sphere of the copied state non-uniformly towards the north pole (i.e. the $|0\rangle$ state). 
As a consequence, a bias on the $\sigma_z$ basis appears when cloning states that lie on the Bloch sphere's equator.
Such footprint can be easily detected in a QKD implementation just by adding a single projective measurement in the $\sigma_z$ basis (for example, detecting $|0\rangle$ states).
To overcome this issue, a desirable property of a PCC machine is for it to be able to prepare clones with no bias but with the same individual fidelity. This can be achieved by  performing the transformation (1) and also its “mirrored” version, that is, one that shrinks the bloch sphere towards the $|1\rangle$ state. 
Several experimental implementations of the PCCM have been performed in discrete variable systems, mainly using photons \cite{sciarrino2005realization,soubusta2008experimental,bartuuvskova2007fiber,xu2008experimental,zhao2005experimental}, NMR systems \cite{chen2007experimental,du2005experimental}, nitrogen-vacancy defect centers in diamond \cite{pan2011solid}, and in continuous variable systems \cite{fiuravsek2001optical,andersen2005unconditional,olivares2006cloning,sabuncu2007experimental}. The use of mirrored PCCM has been introduced in \cite{fiuravsek2003optical,karimipour2002generation,lemr2012experimental}, in the context of assuming partial knowledge about the input state. 

In this work we simulate the mirrored PCC machines using two qubits encoded in the polarization and linear momentum (path) degrees of freedom of a single photon, rather than in two different photons for the two cloned outputs. The interaction is produced in a linear optics setup, via a series of displaced Sagnac polarizing interferometers. We experimentally observe that by alternating randomly between these two operations, Eve can generate a balanced mixture of cloned states and pass unnoticed the test sketched above. In spite of the fact that the polarization-path dual encoding does not produce a physical clone of the transmitted qubits, with this setup we can access to the full range of the cloning parameter $q$, which allows us to study optimal and non-optimal cloning conditions. The asymmetry of the clones is easily tunable by means of a waveplate rotation, without the need of custom beam splitters or imposing additional losses that reduce the overall throughput, as required in two photon 
experiments. Furthermore, the instrinsic stability of the displaced Sagnac 
interferometers does not require any active stabilization, allowing to study the PCC protocol in different regimes. 

Section \ref{sec:exp} of this work presents the experimental setup that implements the phase-covariant cloning machines, and Section \ref{sec:results} is devoted to the results obtained for different experimental conditions. 

\section{Photonic Quantum Simulator}
\label{sec:exp}
The experimental arrangement that implements the PCC machine is shown in Figure \ref{fig:arreglo}. In this scheme, Alice and Bob share a polarization-entangled photonic state. 
Photon pairs are generated by spontaneous parametric downconversion (SPDC) in a BBO type-I nonlinear crystal arrangement \cite{kwiat1999ultrabright,knoll2014remote}, pumped by a 405nm CW laser diode, polarization entanglement is optimized using temporal and spatial compensating birefringent crystals on the pump beam and on the photon pair paths.
Alice's photon is directed into a polarization analysis arrangement where she may perform projective measurements on her photon. Single-photon counting devices detect the incoming photons and send the detections to an FPGA-based coincidence counter. 

\subsection{Linear optics PCC implementation}

\begin{figure}[h!]
    \centering
    \includegraphics[width=0.48\textwidth]{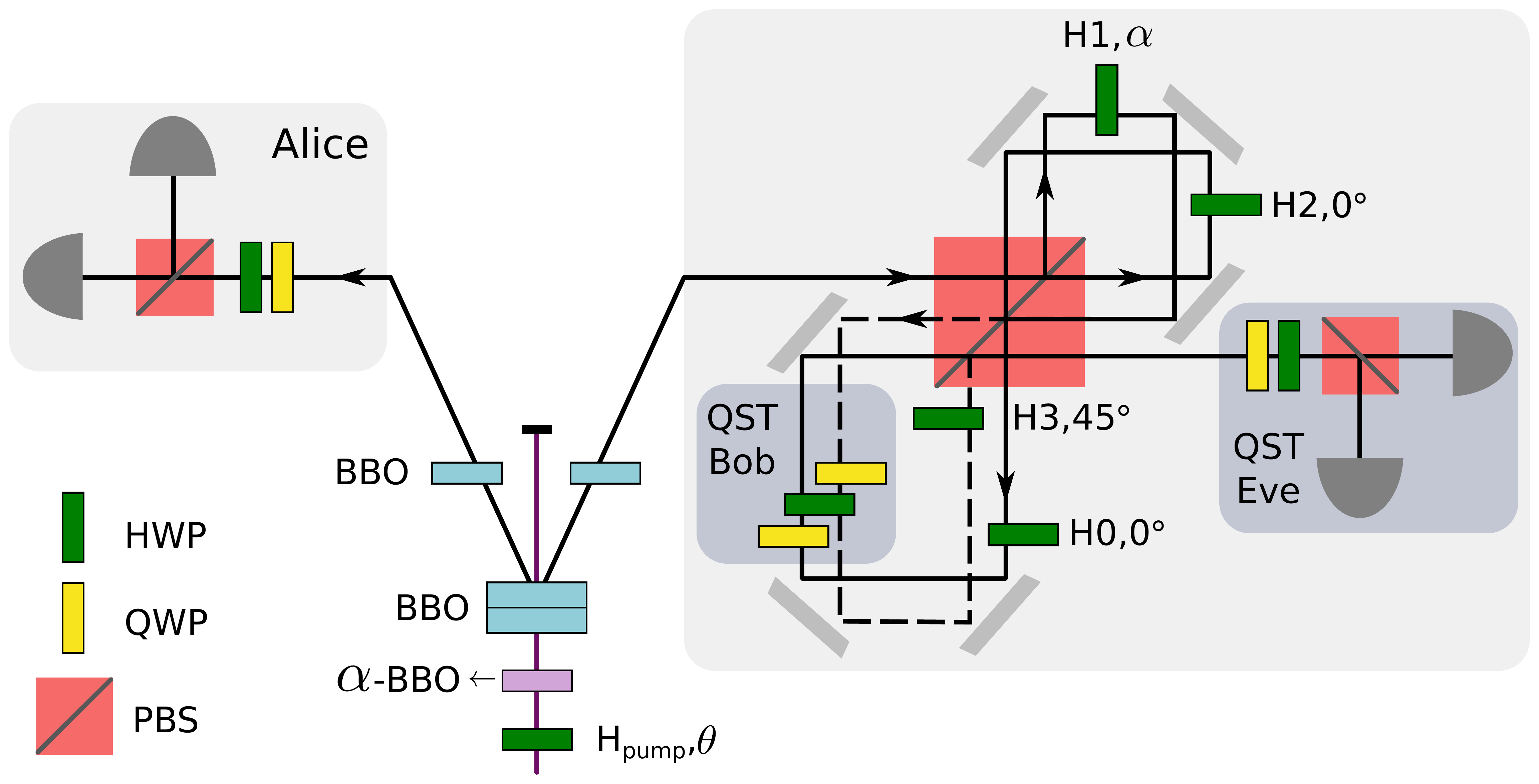}
  \caption{(Color online) Experimental setup for the implementation of the phase covariant cloning machine. The PCC machine is implemented using a displaced Sagnac polarization interferometer: photons enter the interferometer through a PBS, where the $H$ and $V$ polarization components are routed in different directions. If $\alpha=0^{\circ}$ ($H1$) both polarization components are coherently recombined in the PBS and exit the interferometer in the same path. For every other angle of $\alpha$ the $V$ component is transformed into an $H$ polarized photon with probability $q=\sin^{2}(2\alpha)$ and exits the interferometer through path mode $1$, in dashed-lines. Bob may perform quantum state tomography on his polarization qubit before it exits 
the interferometer using a combination of QWPs and a HWP. Both paths are later coherently recombined using a HWP oriented at $45^{\circ}$ (H3) on path $1$ and pass one more time through the polarizing beam splitter. In this way, Eve's qubit is ultimately mapped into a polarization qubit and standard quantum state tomography can be performed using another set of QWP, HWP and a PBS.}
  \label{fig:arreglo}
  \end{figure}

In this quantum simulator, the eavesdropper is encoded on the path qubit of Bob's photon using a displaced Sagnac interferometer based on a polarizing beam splitter (PBS) \cite{almeida2007environment}. An equivalent optical setup has been implemented using calcite beam displacers in \cite{farias2012observation} to study the dynamics of multipartite entanglement.  By associating the horizontal polarization component $H$ with
the state $\ket{0}$ and the vertical polarization $V$ with state $\ket{1}$, and the eavesdropper is
represented by the two path modes of the photon ($0$ and $1$), the PCC transformation is
implemented as follows: an incoming photon in mode $0$ is split into its $H$ and
$V$ components by the PBS. The vertically polarized photons are reflected and propagate
inside the interferometer in the clockwise direction, passing through a
half-wave plate (H1) which transforms the vertical polarization state into 
$\cos(2\alpha)|V\rangle +\sin(2\alpha)|H\rangle$, where $\alpha$ is the physical angle of
the half-wave plate. The horizontal component of this rotated state exits the interferometer, transmitted into mode $1$ 
(dashed-line in Fig.\ref{fig:arreglo}) with probability $q=\sin^{2}(2\alpha)$,
while the vertical component is reflected into mode $0$ with probability $1-q=\cos^{2}(2\alpha)$.

On the other hand, the horizontally polarized photons of the input state propagate through the interferometer in the
counter-clockwise direction and exit through the PBS into mode $0$, with their
polarization state unchanged.
In this way, we obtain the following transformation
\begin{equation}
\begin{aligned}
	&|H\rangle|0\rangle\rightarrow|H\rangle|0\rangle\\
	&|V\rangle|0\rangle\rightarrow\sqrt{1-q}|V\rangle|0\rangle+\sqrt{q}|H\rangle|1\rangle, 
	\label{eq:PCC_exp}
\end{aligned}
\end{equation}
which is a PCC transformation \eqref{eq:PCC} with
$q=\sin^{2}(2\alpha)$. A HWP oriented at $0^{\circ}$ (H2) is placed on the $H$ photons
path to compensate for the optical path difference. The relative path length of the interferometer
is adjusted so that when H1 is oriented at $0^{\circ}$ the polarization
of the input state remains unaltered.

Bob performs standard quantum state tomography (QST) of his polarization state, in coincidence with Alice's detections, using quarter-wave plates on each path, a half-wave plate, and a third passage through the polarizing beam splitter. $H$ polarized photons on mode $1$ are rotated by a HWP
oriented at $45^{\circ}$ (H3) just before passing through the PBS. This operation recombines mode $1$ into
mode $0$ coherently so that both modes exit the interferometer through the same path.
Eve's qubit encoded on Bob's path photon is mapped into a polarization qubit by this operation, 
since photons propagating through mode $0$ are horizontally polarized at the output of the PBS and photons propagating through mode $1$ 
are vertically polarized.
Therefore, Eve can also perform standard QST on her qubit using a QWP, a HWP and a PBS. 

\subsection{Mirrored Device}

We now describe how to obtain the action of a mirrored PCC operation, so that the fidelity for cloning states on the southern hemisphere is now enhanced. This transformation is implemented in the following way: for the input states $\{\ket{1}_B\ket{0}_E, \ket{0}_B\ket{0}_E\}$ we first apply a bit-flip operation on the first qubit, followed by the PCC \eqref{eq:PCC}, and finally another bit-flip on the first qubit, resulting in the following transformation
\begin{equation}
\begin{aligned}
	&|1\rangle_B|0\rangle_E\rightarrow|1\rangle_B|0\rangle_E\\
	&|0\rangle_B|0\rangle_E\rightarrow\sqrt{1-q}|0\rangle_B|0\rangle_E+\sqrt{q}|1\rangle_B|1\rangle_E
	\label{eq:PCCup}
\end{aligned}
\end{equation}
where again $0\leq q\leq1$ controls the asymmetry of the cloning operation. 
It is easy to see that for states on the equator the fidelities of the clones are once again $F_B=\bra{\psi}\rho_B\ket{\psi}=\frac{1}{2}(1+\sqrt{1-q})$ and $F_E=\bra{\psi}\rho_E\ket{\psi}=\frac{1}{2}(1+\sqrt{q})$, independent of the phase $\phi$ of the input state.
This operation shrinks the Bloch sphere of the copied state towards the state $\ket{1}$. By applying randomly these two mirrored operations, Eve can clone the equatorial states with optimized fidelity, while adding no bias on $\sigma_z$.

Interestingly, the mirrored operation can be easily implemented with the same setup by simply rotating half-wave plate H2 instead of H1, which in turn remains fixed at $0^{\circ}$.
In this case, the horizontally polarized photons passing through 
half-wave plate H2 are transformed into 
$\cos(2\alpha')|H\rangle +\sin(2\alpha')|V\rangle$, where $\alpha'$ is the physical angle of
the half-wave plate and we obtain
\begin{equation}
\begin{aligned}
	&|V\rangle|0\rangle\rightarrow|V\rangle|0\rangle\\
	&|H\rangle|0\rangle\rightarrow\sqrt{1-q}|H\rangle|0\rangle+\sqrt{q}|V\rangle|1\rangle, 
	\label{eq:PCCup_exp}
\end{aligned}
\end{equation}
which is equivalent to the mirrored-PCC operation \eqref{eq:PCCup} with
$q=\sin^{2}(2\alpha')$. Throughout the rest of the text we will refer to the machines described in \eqref{eq:PCC_exp} and \eqref{eq:PCCup_exp} as PCC(+)  and PCC(\textendash) respectively.

The action of these mirrored PCC machines can indeed be implemented in two photon experiments, where typically both qubits are encoded on the polarization degree of freedom of two different photons by simply adding half-wave plates before and after the cloning operation, on the signal's qubit path. When rotated at $45^{\circ}$, these HWPs essentially perform a bit-flip operation on the signal qubit, which allows for the mirrored PCC to be implemented as described above. Nevertheless, these experimental implementations require more complicated setups, have low throughput due to multiple coincidence requirements \cite{soubusta2008experimental,soubusta2007several}, and they are less suited to serve as testbeds for different conditions of the cloning protocol.

\section{Results}
\label{sec:results}
With this experimental setup we can prepare and measure an arbitrary qubit state, and control the asymmetry of the cloning operation by simply rotating half-wave plate H1 (or H2). We tested the action of the PCC machines for different experimental conditions: we projected different polarization states on Alice's qubit and performed quantum state tomography on Bob and Eve's qubits. In this way, we obtained the density matrix of both clones and calculated the fidelities for different states and cloning parameters.

\subsection{Equatorial States}

\begin{figure}[h!]
    \centering
    \includegraphics[width=0.45\textwidth]{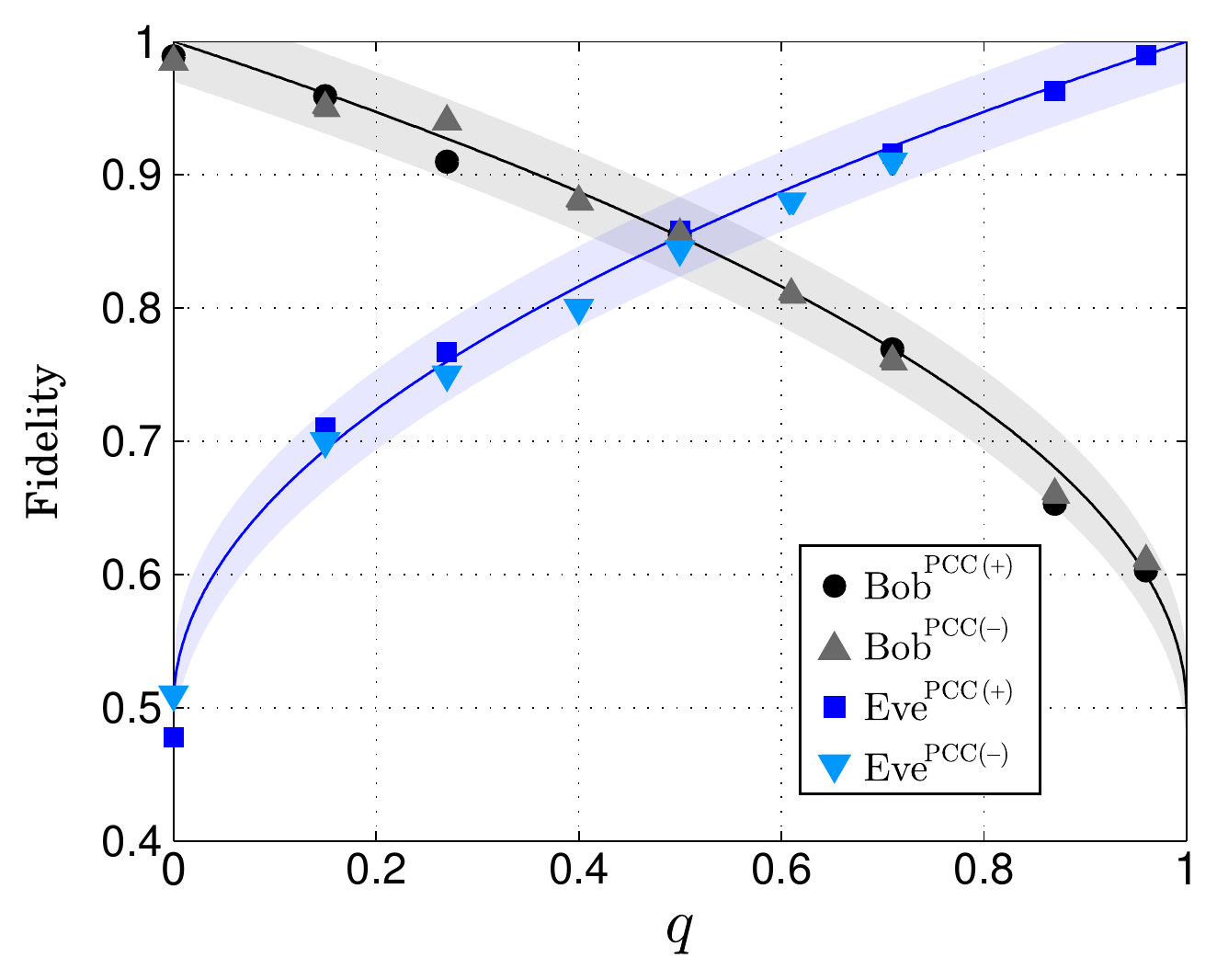}
  \caption{(Color online) Fidelity of the clones with respect to the original state $\ket{D}$, for different cloning parameters $q$, for both PCC (+) and its mirrored operation (\textendash). Filled lines represent the theoretical prediction and the different symbols the experimental data. The shaded area around the theoretical curve accounts for the experimental error.}
  \label{fig:fidq}
  \end{figure}

Figure \ref{fig:fidq} shows the theoretical curves and the experimentally measured fidelities for both Bob and Eve's qubits while Alice's qubit is projected onto state $\ket{D}\bra{D}$, for different values of the cloning parameter $q$ and for both PCC(+)  and PCC(\textendash). The shaded areas represent the experimental error, calculated as the standard deviation of the fidelity values obtained for repeated measurements. As the cloning parameter increases, Bob's fidelity decays to its minimum value $F$=0.5, while Eve's fidelity achieves its maximum when $q$=1. The experimental data shows a good agreement with the theoretical predictions. The fidelity of both cloning machines show the same behavior as a function of the cloning parameter.

\begin{figure}[h!]
    \centering
    \includegraphics[width=0.45\textwidth]{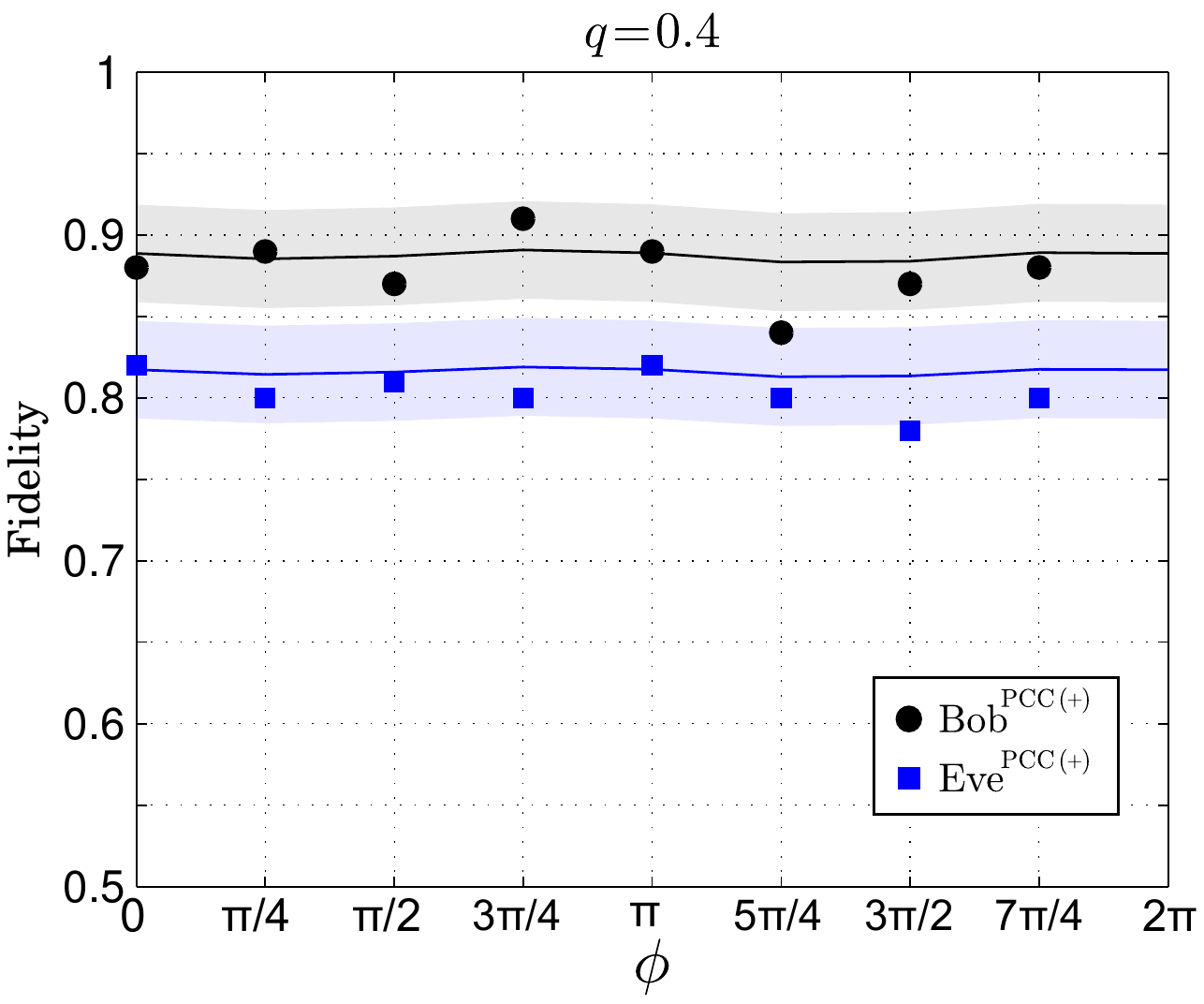}
  \caption{(Color online) Fidelity of the clones for different states on the equator, characterized by the phase $\phi$, for $q$=0.4. The theoretical predictions were calculated numerically based on the entangled input state. The experimental fidelity is shown in black circles for Bob's qubit and blue squares for Eve's. The shaded area represents the experimental error.}
  \label{fig:fidec}
  \end{figure}

A PCC machine clones equally well all states from the equator of the Bloch sphere: by projecting different states on the equator on Alice's side for a particular cloning parameter ($q$=0.4), we obtained the results shown on figure \ref{fig:fidec}. 
The theoretical curves are obtained calculating the fidelity between the ideal equatorial states and the states resulting from projection and cloning processes, applied to the experimentally available input state. Both fidelities for Bob and Eve remain almost constant  within the experimental error (represented by the shaded region), as expected. That is, for states of the form $\ket{\psi}=\frac{1}{\sqrt{2}}(\ket{0}+e^{i\phi}\ket{1})$ the fidelity of the clones is independent of the phase $\phi$, excluding experimental limitations. This particular election of the cloning parameter produces unbalanced fidelities for the two clones.
  
\subsection{Non-Equatorial States}  

\begin{figure}[h!]
    \centering
    \includegraphics[width=0.45\textwidth]{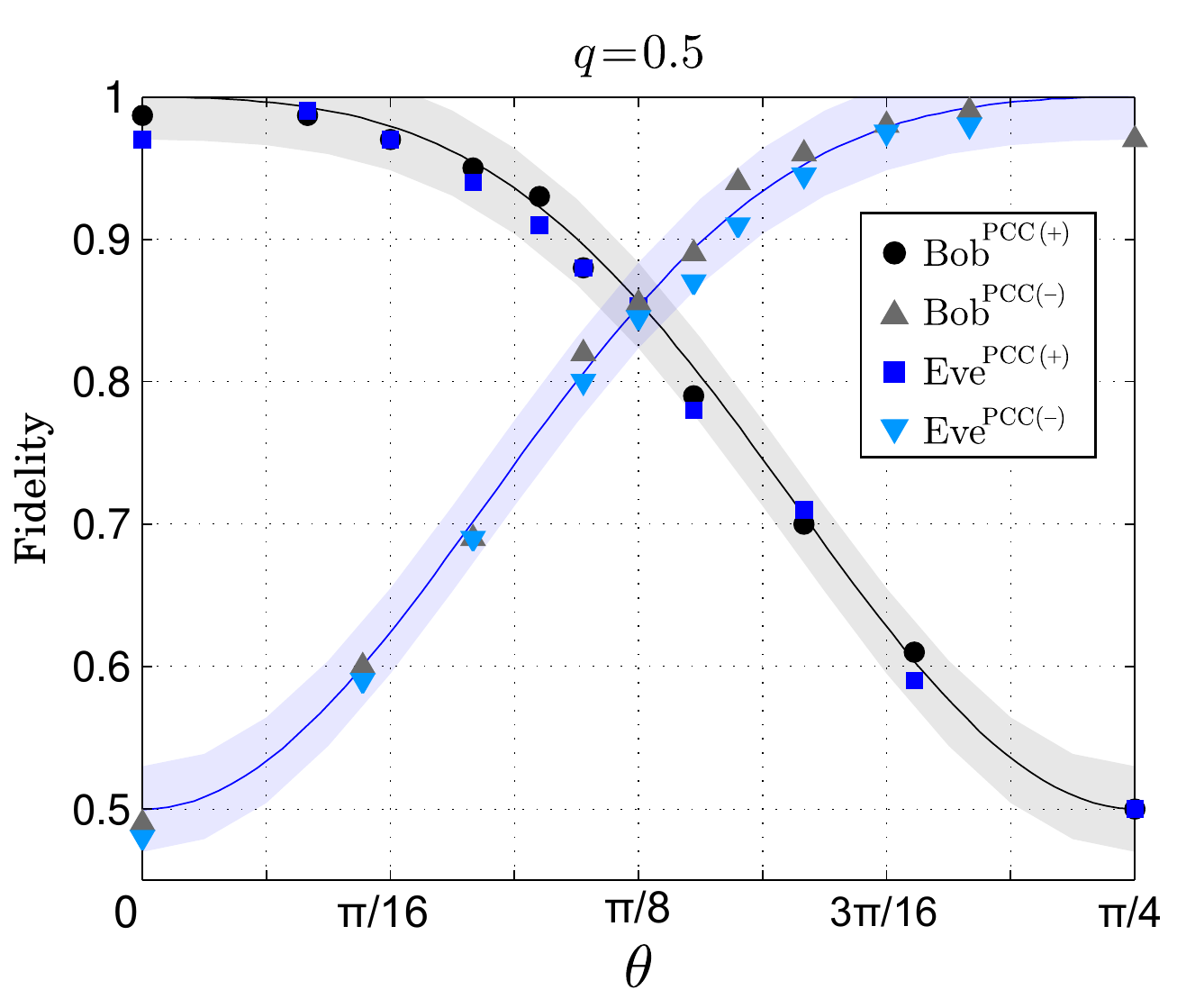}
  \caption{(Color online) Fidelity of the clones for different states of the form $\cos(2\theta)\ket{H}+\sin(2\theta)\ket{V}$, for $q=0.5$. Theoretical curves for the PCC(+) (black line) and PCC(\textendash) operation (blue line) are shown. The different symbols plot the experimental data for both operations and the shaded area represents the experimental error.}
  \label{fig:fidHD}
  \end{figure}

Finally, we projected different states of the form $\ket{\psi}=\cos(2\theta)\ket{H}+\sin(2\theta)\ket{V}$ on Alice's qubit and reconstructed the density matrix for both Bob and Eve's qubits, for $q$=0.5. Figure \ref{fig:fidHD} shows the calculated fidelities for both clones, performing the traditional PCC (black line) and its mirrored operation (blue line). The different symbols represent the experimental data for both operations. Once again, there is a good agreement between the experimental data and the theoretical predictions: PCC(+) clones the state $\ket{H}$ with maximum fidelity ($F$=1) and state $\ket{V}$ with the minimum possible fidelity ($F$=0.5), while the PCC(\textendash) operates in the exact opposite way. Both pairs of curves cross at the balanced superposition state, $\theta=\pi/8$, recovering the symmetric cloning condition with a measured fidelity of $0.852\pm0.005$.

\subsection{Eavesdropping with PCC}
The security limit for the BB84 protocol against incoherent attacks arises from situations where Eve uses a PCC machine \cite{fuchs1997optimal}. This kind of attack can be easily identified if Bob performs measurements on $\sigma_{z}$ (or at least he can project the qubits in one of the two states of the basis), because a side effect of this attack is a bias in the value of $\sigma_{z}$ of the ensemble of received states with respect to the original emitted states ($\sigma_{z}=0$). For a PCC attack the bias in terms of the cloning parameter $q$ is: $\langle\sigma_{z}\rangle=q$.


Such procedure comprises a minor modification to the standard BB84 protocol, given that it can be performed regardless of the states prepared by Alice, and without the need of further classical communication between the parties. In this way, it  can be thought of as an intermediate stage protocol; by evidencing the presence of an eavesdropper, they can eventually devise a suitable strategy to avoid it.

\begin{figure}[h!]
    \centering
    \includegraphics[width=0.45\textwidth]{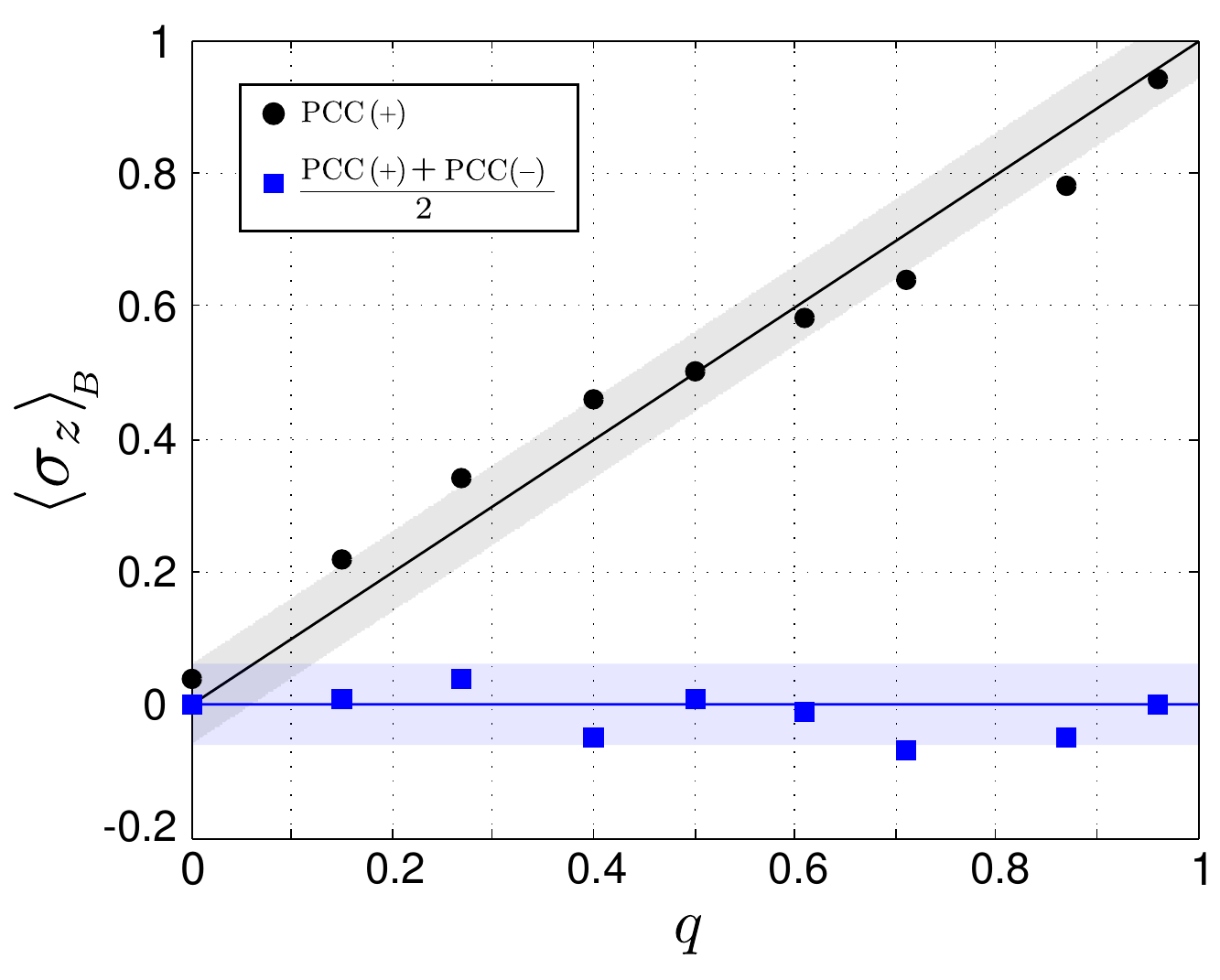}
  \caption{(Color online) Mean value of the $\sigma_z$ component of Bob's reconstructed density matrix for different cloning parameters $q$. The black line and circles correspond to PCC(+) while the blue line and squares represent the averaged values of $\sigma_z$ for an unbiased sequence of PCC(+) and PCC(\textendash) implementations. The shaded area around the theoretical curves accounts for the experimental error.}
  \label{fig:meanH}
  \end{figure}

Nevertheless, the track left by Eve can be erased if she uses a slightly different strategy, that consists on alternating randomly between PCC(+) and PCC(\textendash), thus sending Bob ensembles of cloned states with no bias in $\sigma_z$ but with the same individual fidelity. It should be noted that this strategy does not enable Eve to extract more information than she would have obtained by simply using one of the PCC machines.  
Figure \ref{fig:meanH} shows the effect on the $\sigma_z$ mean value for different cloning parameters $q$ under the action of PPC(+) and the strategy proposed above: while the first protocol adds an increasing bias on the computational basis measurement for increasing values of $q$, the second strategy leaves the $z$ component unbiased, for any strength of the cloning parameter. It can be seen that the experimental implementation of these eavesdropping strategies clearly show the described behaviors.


\section{Conclusions}
\label{sec:conclusion}

We have presented the results of a photonic realization of a protocol that fully simulates a phase-covariant cloning machine, where Alice and Bob share a polarization entangled state, while the eavesdropper is encoded on Bob's path qubit. The cloning algorithm is implemented by  means of a displaced Sagnac interferometer and retardation waveplates.
The versatility of the experimental setup allows us to prepare and measure an arbitrary qubit state, and easily control the asymmetry of the cloning operation in its two mirrored designs. We tested the action of the PCCM and its mirrored version for different experimental conditions, by projecting different polarization states on Alice's qubit, and performing quantum state tomography on Bob and Eve's qubits. 

Regarding an eavesdropping scenario where Eve uses a PCC machine, output states are left unbalanced in the $\sigma_z$ basis after the cloning procedure and could therefore be detected by Bob just by adding a single projective measurement in the $\sigma_z$ basis.
We showed that by alternating between both cloning operations PCC(+) and PCC(\textendash), Eve can mask her presence by sending Bob cloned states with no bias in $\sigma_z$ but with the same individual fidelity.  

\begin{acknowledgement}
 We acknowledge financial support from the Argentine funding agencies CONICET and ANPCyT. 
 We thank Gabriel H. Aguilar for fruitful discussions. 
\end{acknowledgement}


\end{document}